\UseRawInputEncoding
\documentclass[%
 reprint,
 amsmath,amssymb,
 aps,
]{revtex4-1}
\usepackage{graphicx}
\usepackage{dcolumn}
\usepackage{bm}
\usepackage{epsfig}
\usepackage{amsmath}
\usepackage{mathtools}
\usepackage{subfigure}
\usepackage{epstopdf}
\usepackage{textcomp}

\usepackage{url}
\usepackage[colorlinks]{hyperref}

\hypersetup{%
        plainpages=true,
        breaklinks=true,
        hypertexnames=false,
        pageanchor=true,
        colorlinks=true,
        linkcolor={blue},
        citecolor={magenta},
        urlcolor={blue},
        anchorcolor={black}
      }

\usepackage{mathrsfs,bm,amsthm,amsfonts,xspace,float,bookmark}
\usepackage{textcomp}
\usepackage{tabu}
\usepackage{bbold}
\usepackage{xspace}
\usepackage{float}
\usepackage[inline]{enumitem}

\renewcommand{\thesection}{\Roman{section}}

\newcommand{\kb}[1]{\langle#1\rangle}
\newcommand{\unite}[1]{\,\rm#1}
\usepackage[normalem]{ulem}

\DeclareMathOperator{\Tr}{Tr}

\begin{document}

\author{Yong Lu$^{1}$}
\email[e-mail:]{yongl@chalmers.se}

\author{Andreas Bengtsson$^{1}$}

\author{Jonathan J.Burnett$^{1,2}$}
\author{Baladitya Suri$^{1,3}$}
\author{Sankar Raman Sathyamoorthy$^{1}$}
\author{Hampus Renberg Nilsson$^{1}$}
\author{Marco Scigliuzzo$^{1}$}
\author{Jonas Bylander$^{1}$}
\author{G\"{o}ran Johansson$^{1}$}

\author{Per Delsing$^{1}$}
\email[e-mail:]{per.delsing@chalmers.se}

\address{$^1$Microtechnology and Nanoscience, Chalmers University of Technology, SE-412 96, G\"{o}teborg, Sweden
\\$^2$National Physical Laboratory, Hampton road, Teddington, TW11 0LW, UK
\\$^3$Indian Institute of Science, Department of Instrumentation and Applied Physics,Bangalore 560012, India
}

\title{Quantum efficiency, purity and stability of a tunable, narrowband microwave single-photon source}
\begin{abstract}
We demonstrate an on-demand source of microwave single photons with 71--99\% intrinsic quantum efficiency. The source is narrowband (300\unite{kHz}) and tuneable over a 600 MHz range around 5.2 GHz. Such a device is an important element in numerous quantum technologies and applications. The device consists of a superconducting transmon qubit coupled to the open end of a transmission line. A $\pi$-pulse excites the qubit, which subsequently rapidly emits a single photon into the transmission line. A cancellation pulse then suppresses the reflected $\pi$-pulse by 33.5 dB, resulting in 0.005 photons leaking into the photon emission channel. We verify strong antibunching of the emitted photon field and determine its Wigner function. Non-radiative decay and $1/f$ flux noise both affect the quantum efficiency. We also study the device stability over time and identify uncorrelated discrete jumps of the pure dephasing rate at different qubit frequencies on a time scale of hours, which we attribute to independent two-level system defects in the device dielectrics, dispersively coupled to the qubit.
\end{abstract}
\maketitle

\section{INTRODUCTION}
\label{sec1}
The single photon---the fundamental excitation of the electromagnetic field---plays a key role in quantum physics and can find practical application in quantum sensing \cite{degen2017quantum}, communication~\cite{kimble2008quantum} and computing~\cite{knill2001scheme,kok2007linear,zhong2020quantum}.
Recently, considerable progress has been made in the generation of optical photons, e.g.\@ by using quantum dots~\cite{somaschi2016near,senellart2017high, schweickert2018demand}. However, in the microwave domain, the much smaller photon energy introduces many constrains for the realisation of single-photon sources; for instance, operation at millikelvin temperatures is necessary to avoid thermal generation of photons.
Narrowband microwave single photons are essential for precise interactions with circuits exhibiting a shaped energy structure, such as coplanar resonators~\cite{barends2008quasiparticle}, three-dimensional cavities~\cite{reagor2013reaching} and acoustic-wave resonators~\cite{maccabe2020nano,chu2018creation}, which can be used as quantum memories.

Superconducting quantum circuits are suitable for the implementation of on-demand microwave photon sources. So far, several different methods have been used.
The first method is based on a qubit coupled to a resonator
~\cite{houck2007generating,lang2013,pechal2014microwave}, where the source bandwidth is limited by the linewidth of the resonator.
Secondly, in Refs.~\cite{leppakangas2015antibunched, grimm2019bright, rolland2019antibunched}, single photons are generated due to inelastic Cooper-pair tunnelling. This type of source has a high emission rate, but it cannot generate a superposition of vacuum and a single-photon Fock state.
Thirdly,  a single-photon generator based on emission from a qubit into a waveguide requires proper engineering of the asymmetric couplings to the control and emission channels~\cite{lindkvist2014scattering,peng2016tuneable,pechal2016superconducting,zhou2020tunable}. 
Finally, shaped single photons emitted from a qubit located near the end of a transmission line with a tunable-impedance termination~\cite{sathyamoorthy2016simple} were demonstrated in experiment~\cite{forn2017demand}.
None of these experiments included a thorough study of the photon leakage of the excitation pulse from the control to the emission channel, which affects the purity of the single-photon.

In this work, we implement a different proposal from Ref.~\cite{sathyamoorthy2016simple}: a frequency-tunable qubit is capacitively coupled to the end of an open transmission line \cite{hoi2015probing,lin2020deterministic}. Only a single channel exists in our system, so that the qubit, excited by a $\pi$-pulse, can only release a single photon back to the input. We cancel the $\pi$-pulse, after its interaction with the qubit, by interfering it with another, phase-shifted pulse and show a photon leakage $0.5\%$ of a photon from the excitation pulse. The intrinsic quantum efficiency of our single-photon source is 71--99\% over a tuneable frequency range of $600\unite{MHz}$ around $5.2\unite{GHz}$, which is about 1600 times larger than the single-photon linewidth (300\unite{kHz}). This bandwidth is more than 20 times narrower than that of the tuneable microwave single-photon sources reported in Refs.~\cite{leppakangas2015antibunched, grimm2019bright, rolland2019antibunched,peng2016tuneable,pechal2016superconducting,zhou2020tunable,forn2017demand}.

Importantly, the intrinsic quantum efficiency---the fidelity only due to the emitter coherence---can be limited by both the pure dephasing rate and the non-radiative decay rate of the emitter. It is important to understand the noise mechanisms determining these rates in order to make further improvements.
We systematically study the limitation of the intrinsic quantum efficiency and the temporal fluctuations of the single photon source over 136 hours. The result shows that both nonradiative decay and $1/f$ flux noise can affect the quantum efficiency from different types of two-level fluctuators. In addition, we also characterise the fluctuations of the pure dephasing rate due to dispersively coupled two-level system defects with a narrow linewidth, which can lead to a decrease of the quantum efficiency by up to $60\%$.

\section{Results}
\begin{figure}
\includegraphics[width=1\linewidth]{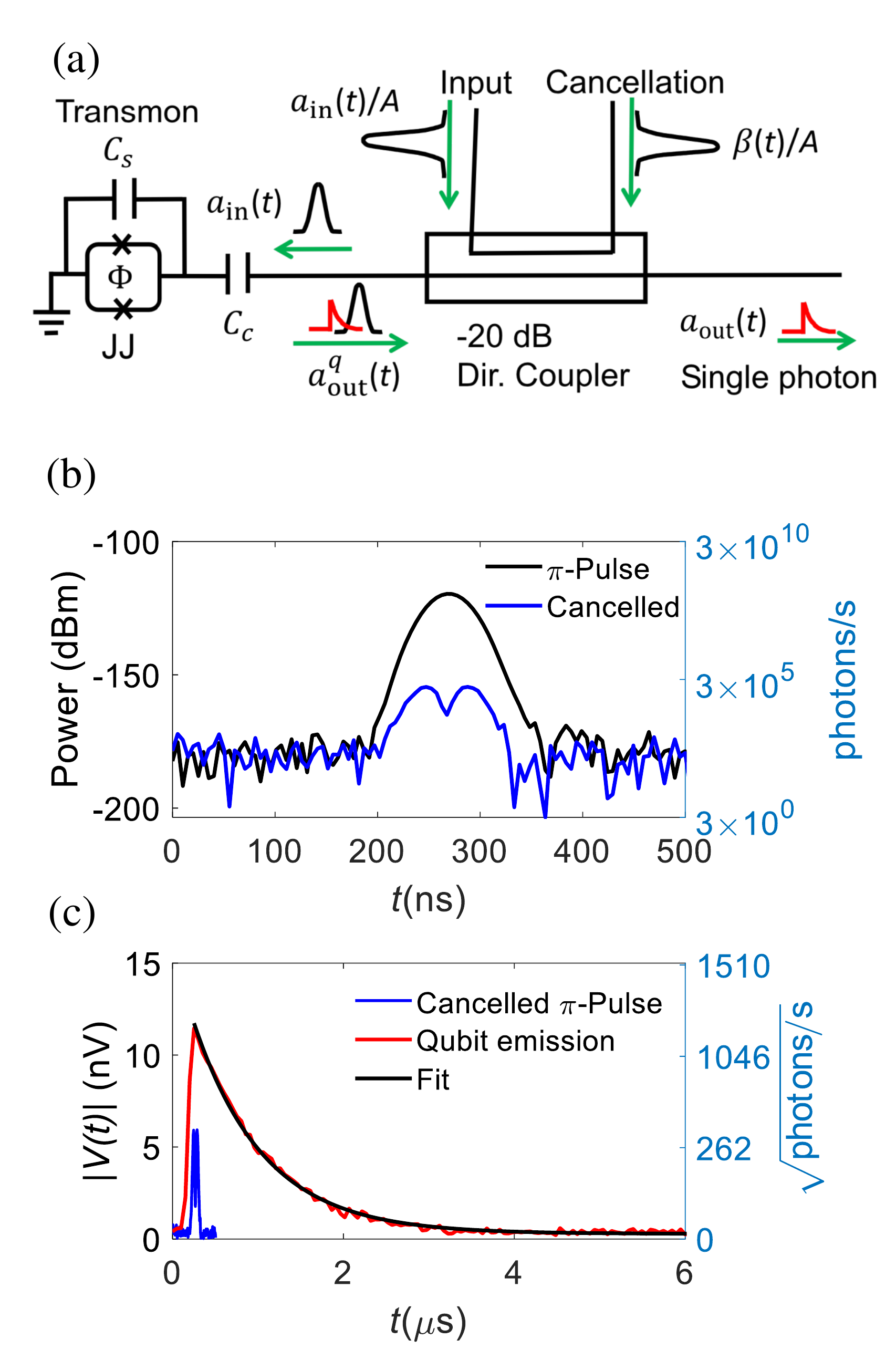}
\caption{Pulse cancellation.
(a) Schematic for generating single photons using pulse cancellation. A flux-tunable transmon-type superconducting qubit (artificial atom) capacitively coupled to the end of an open transmission line with a -20~dB directional coupler connected to the transmission line. $C_{\rm{s}}$ and $C_{\rm{c}}$ represent the shunt capacitance for the qubit and the coupling capacitance between the qubit and the transmission line, respectively. $\Phi$ is the external magnetic flux threading the SQUID (Superconducting QUantum Interference Device) loop and JJ denotes the Josephson junctions.
(b) Comparison of a $\pi$-pulse with and without the cancellation when the qubit is tuned away. The input pulse is suppressed by -33.5dB with a cancellation pulse with $5.12\times10^5$ averages.
(c) Comparison between the cancelled $\pi$-pulse from (b) and the photon emission by the qubit (red line) after a $\pi/2$-pulse with the pulse cancellation on. The red line is a fit to an exponential decay to extract $\Gamma_2/(2\pi)=193\pm4\unite{kHz}$.
}
\label{cancellation}
\end{figure}

{\bf{Experimental setup and procedure for single photon emission.}} Our device consists of a magnetic-flux-tunable Xmon-type transmon qubit, capacitively coupled to the open end of a one-dimensional coplanar-waveguide transmission line. This zero-current boundary condition behaves as a mirror for the incoming microwave radiation.  The corresponding simplified circuit diagram is shown in Fig~\ref{cancellation}(a). An asymmetric beam splitter, implemented by a 20 dB directional coupler, is connected to the sample to provide channels for qubit excitation and pulse cancellation.  The circuit is made of aluminium on a silicon substrate, and is fabricated with a standard lithography process \cite{burnett2019decoherence}. The sample is characterized at $T=10\unite{mK}$ with its parameters shown in Table~\ref{tab:1}.
\begin{table}
 \caption{Device parameters. The qubit parameters are obtained by single- and two-tone spectroscopy from the reflection coefficient measurements (see more details in the Methods section). The qubit frequency $\omega_{01}(\Phi)$ depends on the external flux $\Phi$ and we define $\omega_{01,1}=\omega_{01}(\Phi=0)$. $\alpha$ is the qubit anharmonicity, $\Gamma_{\rm{r}}$ and $\Gamma_{\rm{2}}$ are the radiative decay rate and the decoherence rate of the qubit. The error bars within parenthesis are two standard deviations.  } \label{tab:1}
  \centering
\begin{tabular*}{\columnwidth}{  @{\extracolsep{\fill}} c c c c c c  @{} }
\hline
\hline
  $\alpha$ & $\omega_{01,1}/2\pi$ &$\Gamma_{\rm{r}}/2\pi$&$\Gamma_2/2\pi$\\
   ~\rm{GHz}&~\rm{GHz}&~\rm{kHz}&~\rm{kHz}\\
\hline
 0.251 & 5.510&270~(1)&188~(1)\\
\hline
\hline
\end{tabular*}

\end{table}

As shown in Fig.~\ref{cancellation}(a), we send a pulse to the input port of the directional coupler with the amplitude $a_{\rm{in}}(t)/A$, where $A=0.1$ is the attenuation from the -20\unite{dB} directional coupler. Then, $a_{\rm{in}}(t)$ is the corresponding amplitude of the pulse at the qubit.  The output field at the qubit, using the standard input-output relation, is $a_{\rm{out}}^{\rm{q}}(t)=a_{\rm{in}}(t)-i\sqrt{\Gamma_{\rm{r}}}\sigma_-(t)$ \cite{sathyamoorthy2016simple,koshino2012control}, where $\sigma_-(t)$ is the emission operator of the qubit. By adding another pulse $\beta(t)/A$ to the cancellation port of the directional coupler, we have $a_{\rm{out}}(t)=a_{\rm{out}}^{\rm{q}}(t)+\beta(t)$ at the output of the directional coupler. When $\beta(t)=-a_{\rm{in}}(t)$, we have $a_{\rm{out}}(t)=-i\sqrt{\Gamma_{\rm{r}}}\sigma_-(t)$ (the small red pulse). This means we obtain a single photon if $a_{\rm{in}}(t)$ is a $\pi$-pulse, and a superposition of vacuum and a single-photon Fock state if $a_{\rm{in}}(t)$ is a $\pi/2$-pulse.

\begin{figure}
\includegraphics[width=\linewidth]{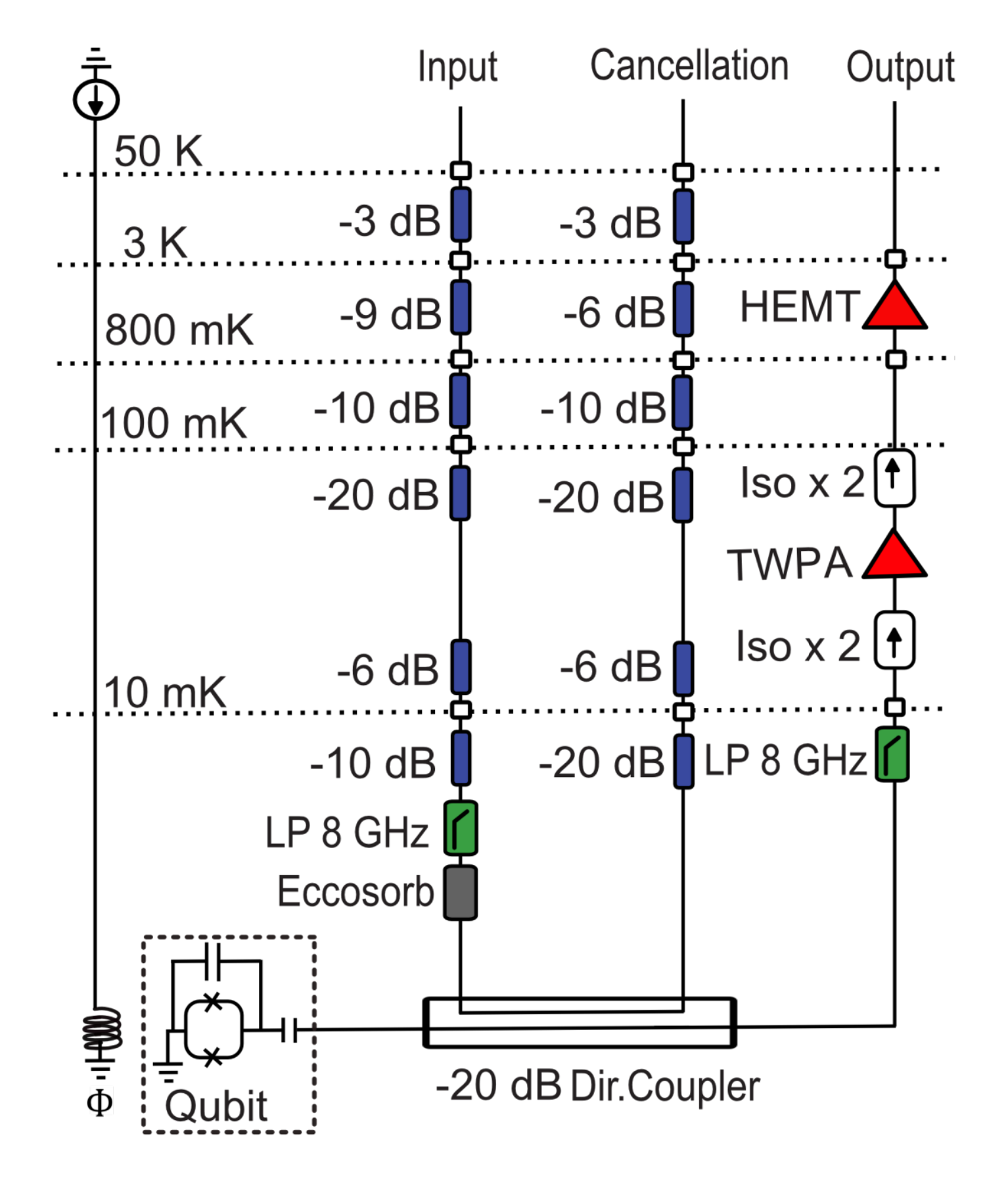}
\caption{The measurement setup. LP, Iso, HEMT, and TWPA denote low-pass filters, isolators, a
high electron mobility transistor amplifier, a traveling wave parametric amplifier.
}
\label{setup}
\end{figure}

We adjust the external flux to zero ($\Phi=0$) so that the qubit reaches its highest frequency $\omega_{01}$. We then send a calibrated 50\unite{ns} Gaussian pulse on resonance with the qubit, so that it acts as a $\pi$-pulse. We measure the output field using a traveling-wave parametric amplifier (TWPA) \cite{macklin2015near} followed by a
high electron mobility transistor amplifier (HEMT) [Fig.\ref{setup}]. Both quadratures of the signal output from the directional coupler, with and without the cancellation, are amplified and recorded by a digitizer (not shown) as a voltage $V(t)=I(t)+iQ(t)$. The voltage is then normalized by the system gain from the on-resonance Mollow triplet \cite{lu2021characterizing,lu2021propagating,astafiev2010resonance}. After averaging, the corresponding photon number at the qubit is defined as

\begin{equation}
n=\frac{1}{2Z_0\hbar\omega_{01}}\int_{t_0}^{t_1} \left( |\langle V(t)\rangle|^2-|\langle V_{N}\rangle|^2 \right) \,\mathrm{d}t
\label{photonnumber}
\end{equation}
where $t_0$ and $t_1$ denote when the signal starts and ends, respectively. $\langle V_{N}\rangle$ is the averaged system voltage noise and $Z_0\approx 50\unite{\Omega}$ is the waveguide impedance.

Figure~\ref{cancellation}(b) shows the power of the input $\pi$-pulse as a function of time. The black line indicates the power of the input pulse at the sample after the gain calibration when the qubit is tuned away, while the blue one corresponds to the residual pulse after cancellation. The result shows a 33.5 dB suppression of a $\pi$-pulse in power due to the cancellation, resulting in a photon leakage of $n_{\rm{leak}}^{\rm{meas}}=0.0049$, according to Eq.~(\ref{photonnumber}). In Fig.~\ref{cancellation}(c), we also measure the coherent emission (red line) from the qubit decay after a $\pi/2$-pulse and fit the data to an exponential curve (black) with a decay rate $\Gamma_2/(2\pi)=193\pm4\unite{kHz}$. By taking the integral over time with Eq. (\ref{photonnumber}), we obtain the photon numbers $n_{\rm{q}}^{\rm{meas}}\approx0.173$ for the qubit emission. This agrees well with the formula $\Gamma_{\rm{r}}/8\Gamma_{2}=0.1795$ derived below. We notice that $n_{\rm{q}}^{\rm{meas}}$ is less than 0.5 since we just measure the coherent part of the qubit emission.

The leakage from the excitation pulse can also be estimated without calibrating the system gain as follows. The driven qubit generates a voltage amplitude of $V_{\rm{q}}(t)=i2\omega_{01}Z_0C_{\rm{c}}d\sigma_-(t)$\cite{peng2016tuneable}, where $d$ is the qubit dipole moment, and $C_{\rm{c}}$ represents the coupling capacitance between the qubit and the transmission line. The radiative decay rate is given by $\Gamma_{\rm{r}}=S_{\rm{v}}(\omega)(C_{\rm{c}}d)^2/\hbar^2$ with $S_{\rm{v}}(\omega)=2\hbar\omega_{01} Z_0$ being the spectral density of the voltage quantum noise in the transmission line where we ignore the effect from the thermal noise inside the waveguide since $\hbar\omega_{01}\gg k_BT$. Therefore, the corresponding emission power from the qubit is $|V_{\rm{q}}(t)|^2/(2Z_0)=\hbar\omega_{01}\Gamma_{\rm{r}}|\sigma_-(0)|^2e^{-2\Gamma_2t}$ where we have $\sigma_-(t)=\sigma_-(0)e^{-\Gamma_2t}$. By taking the integral over time, the photon number is $n_{\rm{q}}={\Gamma_{\rm{r}}}/(2\Gamma_2)|\sigma_-(0)|^2={\Gamma_{\rm{r}}}/{8\Gamma_2}$. Combining the values of $\Gamma_{\rm{r}}$ and $\Gamma_{\rm{2}}$ in Table~\ref{tab:1}, the leakage from the $\pi$-pulse is $n_{\rm{leak}}=n_{\rm{leak}}^{\rm{meas}}/n_{\rm{q}}^{\rm{meas}}\times n_{\rm{q}}\approx 0.005$. In reality, $|\sigma_-(0)|<0.5$ due to the small emission during a $\pi/2$-pulse. Here, we ignore this since our pulse length is much shorter than the qubit lifetime. We emphasize that the amplitude of the cancelled pulse in Fig.~\ref{cancellation}(b) and (c) was minimized by adjusting the amplitude of the cancellation pulse, the phase difference between the input and the cancellation pulse and compensating the time delay between these two pulses. Compared to directly measuring the qubit emission power after a $\pi$-pulse, we take an advantage of the coherent emission after a $\pi/2$-pulse so that the system noise can be averaged out.

\begin{figure}
\includegraphics[width=1\linewidth]{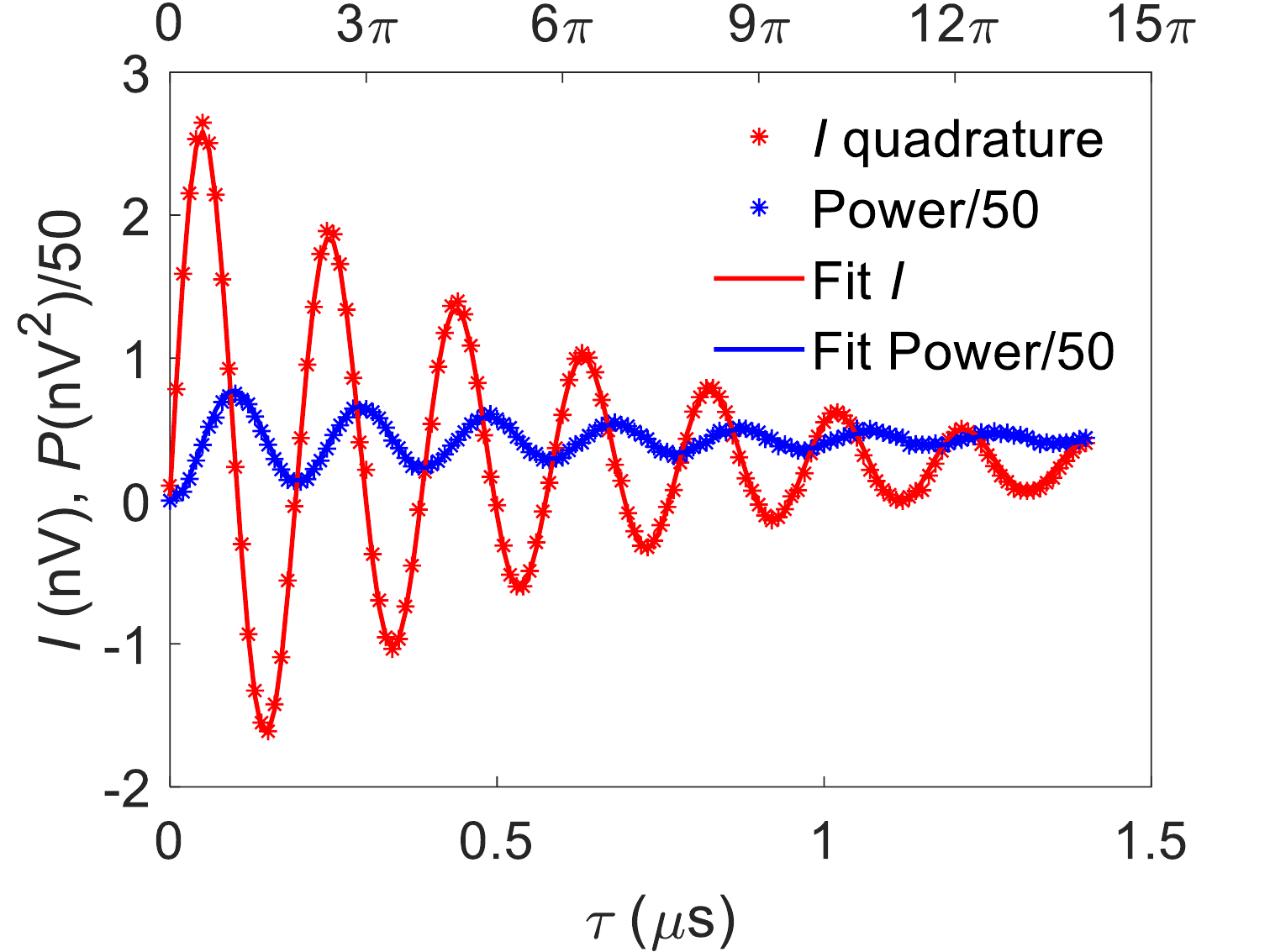}
\caption{Quadrature and power oscillations of emitted radiation from the driven qubit vs pulse length $\tau$. Red stars represent the measured $I$ quadrature amplitude, while blue stars correspond to the emitted power $\langle P\rangle=\kb{(I^2+Q^2)}$. Both traces are fitted to a sinusoid function with an exponential-decay envelope, simultaneously. The extracted decay rate is $2\pi*629\unite{kHz}$. Moreover, the phase of the two fitting curves is offset by $\pi/2$, which rules out a coherent state and provides evidence for single-photon emission. The top axis indicates the angle of the qubit-state rotation on the Bloch sphere. The external flux is $\Phi=0$.
}
\label{oscillation}
\end{figure}

{\bf{Qubit Operation.}} Next we vary the pulse length $\tau$, and measure the integral of free-decay traces such as the one in Fig.~\ref{cancellation}(c), normalized to the number of points in the trace. In order to maximize the signal, we digitally rotate the integrated value into the $I$ quadrature. Meanwhile, we also record the second moment of the emitted field which corresponds to the emitted power $\langle P\rangle=\kb{(I^2+Q^2)}$. Figure~\ref{oscillation} shows the Rabi oscillations of $\kb{I}$ and $\kb{P}$ with pulse lengths up to $1.4\unite{\mu s}$. The signal is averaged over $1.28*10^4$ repetitions. The background offset from the system noise is removed from each data point of the power oscillation. The clear oscillatory pattern in the figure is a manifestation of the coherence of photons emitted by the qubit. By solving the Bloch equations we obtain
%
%
\begin{eqnarray}
\kb{\sigma_y}&=&\frac{\Omega}{\Omega^2+\Gamma_1\Gamma_2}[\Gamma_1+{e^{-\Gamma_{\rm{s}}\tau}}{\sqrt{\Gamma_1^2+B_1^2}}\sin(\Omega_{\rm{m}}\tau-\theta_1)]\nonumber\\
\kb{\sigma_z}&=&\frac{-\Gamma_1\Gamma_2-\Omega^2{e^{-\Gamma_{\rm{s}}\tau}}{\sqrt{1+B_2^2}}\sin(\Omega_{\rm{m}}\tau+\theta_2)}{\Omega^2+\Gamma_1\Gamma_2},
\label{oscillationth1}
\end{eqnarray}
where $\Gamma_{\rm{1}}$ is the relaxation rate of the qubit and $\Omega$ is the Rabi frequency, $\Gamma_{\rm{s}}=(\Gamma_1+\Gamma_2)/2$, $\Omega_{\rm{m}}=\sqrt{\Omega^2-(\Gamma_1-\Gamma_2)^2/4}$, $B_1=\Omega_{\rm{m}}-({\Gamma_1^2-\Gamma_2^2})/({4\Omega_{\rm{m}}})$, $B_2={\Gamma_{\rm{s}}}/{\Omega_{\rm{m}}}=\cot\theta_2$ and ${\Gamma_1}/{\Omega_{\rm{m}}}\approx\tan{\theta_1}$. Since $\kb{I}\propto\kb{\sigma_y}$ and $\kb{P}\propto1+\kb{\sigma_z}$, we take Eq.~(\ref{oscillationth1}) to fit the data to obtain $\Gamma_{\rm{s}}/2\pi=316\pm6\unite{kHz}$ and $\theta_2+\theta_1=(0.498\pm0.004)\pi$. The phase difference indicates that the measured radiation is not from a coherent state in which the power and amplitude would oscillate in phase.

%
%

\begin{figure}
\includegraphics[width=1\linewidth]{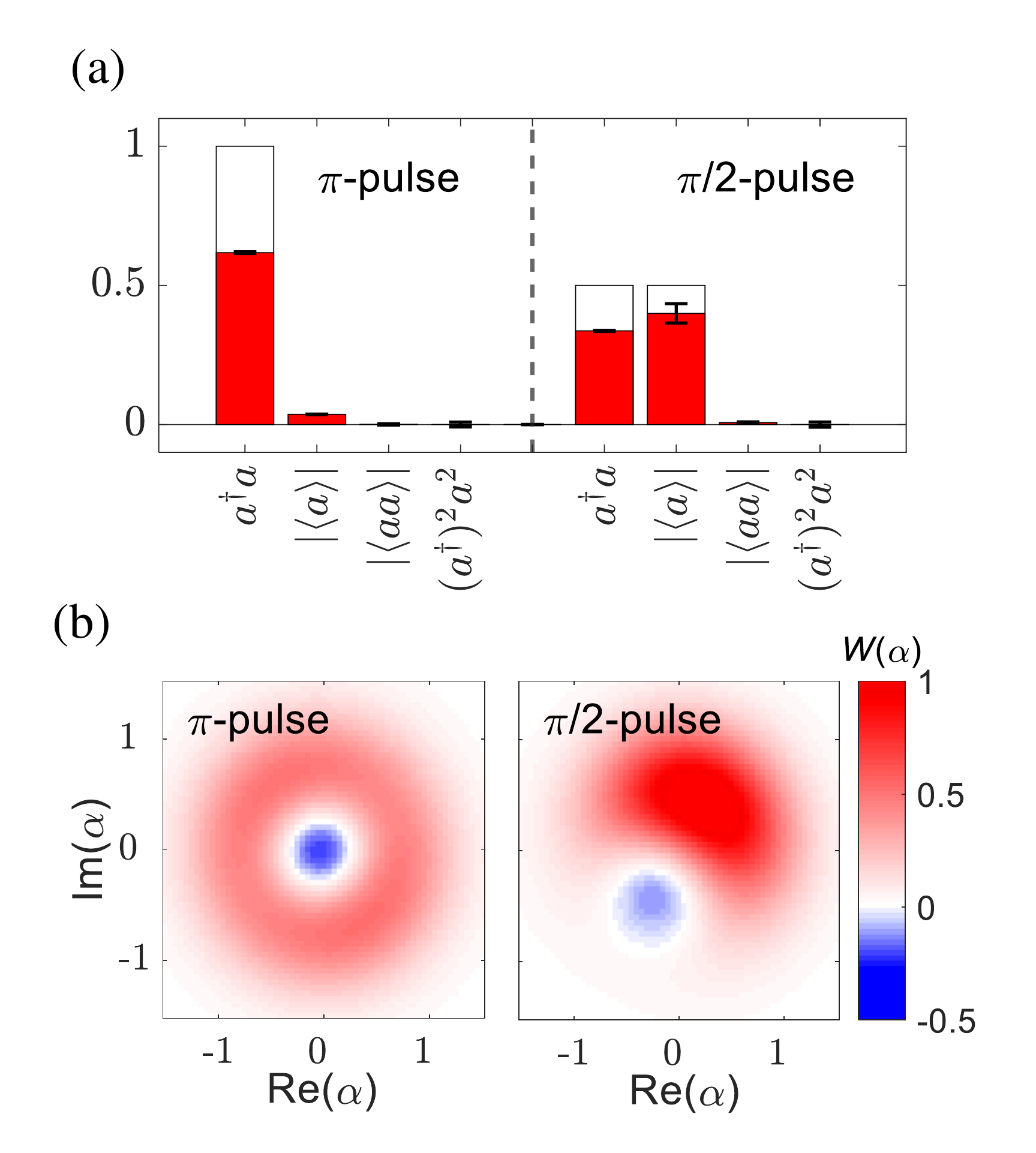}
\caption{(a) The bar chart shows a comparison between the experiment (red) and theory (white) for the moments of a single-photon state $|1\rangle$ after a $\pi$-pulse pulse and a superposition state $(|0\rangle +|1\rangle)/\sqrt{2}$ after a $\pi/2$-pulse. (b) Wigner functions corresponding to the moments obtained from the experiment in (a) using a maximum likelihood method~\cite{James2001,eichler2012characterizing}.
}
\label{moment}
\end{figure}

To demonstrate that our device indeed is a single-photon source we extract the second order correlation function $g^{(2)}(0)$ and we reconstruct the Wigner function $W(\alpha)$~\cite{Cahill1969}. We send either a $\pi$-pulse or a $\pi/2$-pulse to excite the qubit. With an appropriate mode-matching filter with an exponential decay, we obtain the quadrature histograms of the measured single-shot voltages normalized by the gain value. The single-shot measurement is repeated up to $2.56*10^7$ times. By then subtracting the reference values measured in the absence of the pulse, as outlined for example in Ref.~\cite{eichler2012characterizing}, we extract the moments of the photon mode $a$. Figure.~\ref{moment}(a) shows the moments $|\langle a\rangle|$,
$\langle a^\dag a\rangle$ and $\langle (a^\dag)^2 a^2\rangle$ of the qubit emission after a $\pi$-pulse and a $\pi/2$-pulse, respectively. The first and second-order moments are $0.036\pm0.001$ and $0.618\pm0.003$ for a $\pi$-pulse, and $0.399\pm0.035$ and $0.337\pm0.002$ for a $\pi/2$-pulse. The second order of moments shows that the overall quantum efficiencies at the maximum qubit frequency are $61.8\%$ for a single-Fock state $|1\rangle$ after a $\pi$-pulse, and $67.4\%$ for a superposition state $(|0\rangle+|1\rangle)/\sqrt{2}$ after a $\pi/2$-pulse. In our case, the maximum photon number is just one so that we only need to consider up to the fourth order of the moments corresponding to two photons.

\begin{figure}
\includegraphics[width=1\linewidth]{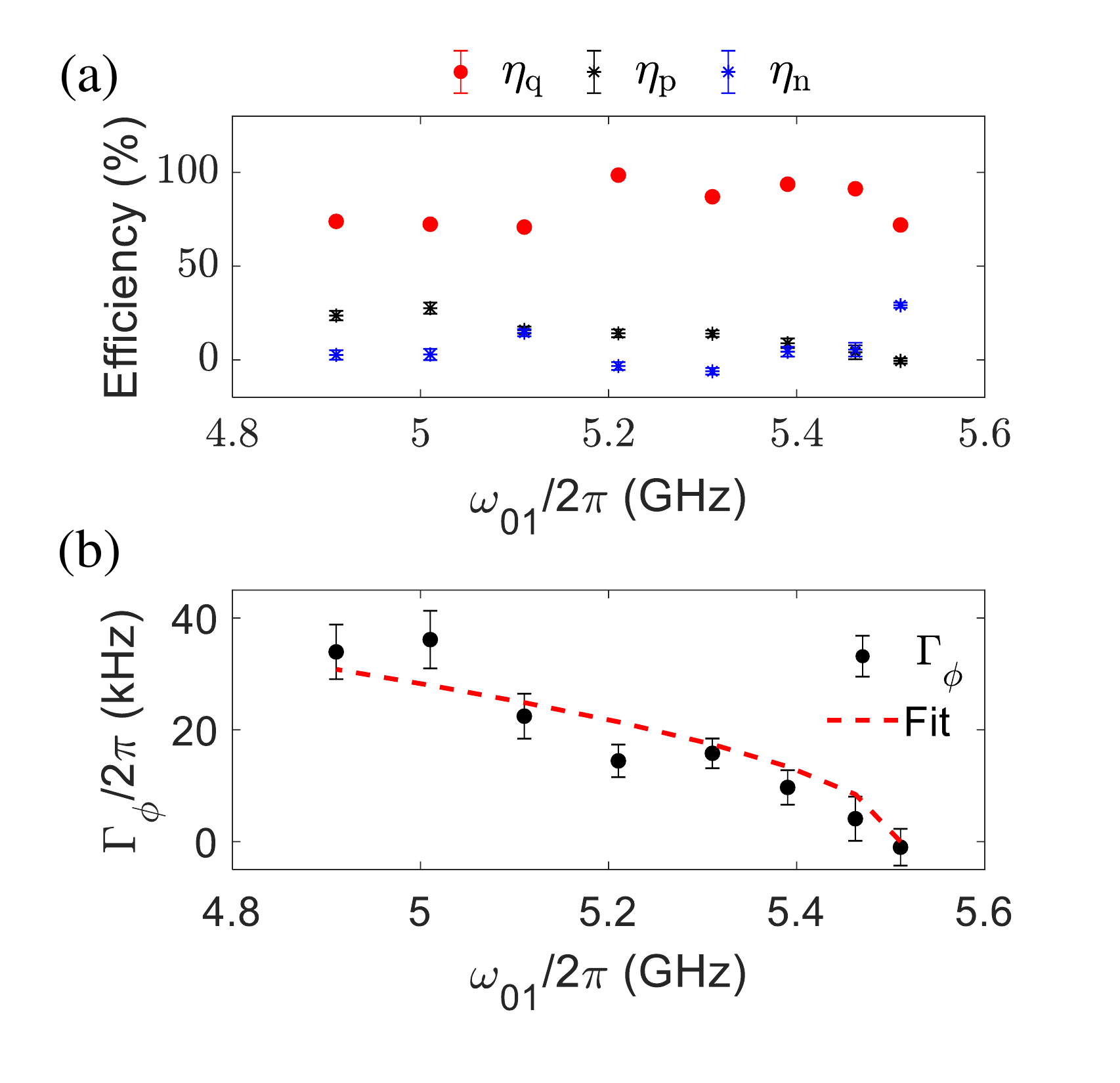}
\caption{
(a) The intrinsic quantum efficiency $\eta_{\rm{q}}$ for our single photon source over the 600\unite{MHz} tunable range. The efficiency is limited by the pure dephasing rate and the non-radiative decay rate of the qubit. These two factors reduce the efficiency by $\eta_{\rm{p}}$ and $\eta_{\rm{n}}$ respectively, where we have $\eta_{\rm{q}}+\eta_{\rm{p}}+\eta_{\rm{n}}=1$.
(b) Pure dephasing rate $\Gamma_{\phi}$ as a function of the qubit frequency. In all panels, the error bars are two standard deviations.
}
\label{efficiency}
\end{figure}

The moments we extract differ from the theoretically expected $\langle a^\dag a\rangle=1$ for the Fock state and $|\langle a\rangle|=0.5$ for the superposition state.
The numerical result from simulating the dynamics of the qubit by using QuTip~\cite{johansson2012qutip} shows that the population of the first excited level of our qubit is given by the density matrix element $\rho_{11}=0.93$ after a 50\unite{ns} $\pi$-pulse, and $|\sigma_-|=0.44$ after a 50 ns $\pi/2$-pulse. The normalized filter for the mode matching is $f(t)=\sqrt{\Gamma_1}e^{-{\Gamma_1}/{2}t}$, leading to $\langle a^\dag a\rangle=\Gamma_{\rm{r}}/\Gamma_1$ and
$|\langle a\rangle|=2\sqrt{\Gamma_{\rm{r}}\Gamma_1}/(2\Gamma_2+\Gamma_1)$. In summary, we have $\langle a^\dag a\rangle=0.93*\Gamma_{\rm{r}}/\Gamma_1$ and $\langle a\rangle=0.44*2\sqrt{\Gamma_{\rm{r}}\Gamma_1}/(2\Gamma_2+\Gamma_1)$. Combining the decay rates in Table~\ref{tab:1} and assuming that the pure dephasing rate is zero [Fig.~\ref{efficiency}(b)] at the maximum qubit frequency ($\Phi=0$), we get $\langle a^\dag a\rangle=0.67$ and $|\langle a\rangle|=0.36$, which are close to our measured results. From this discussion, we can conclude that the non-radiative decay is the main factor that limits the quantum efficiency of our single-photon source at the flux sweet spot, and the overall quantum efficiency are limited by both the imperfect qubit excitation and the qubit coherence.

Of particular interest is the normalized zero-time-delay intensity correlation function $g^{(2)}(0)=\langle (a^\dag)^2 a^2\rangle/\langle a^\dag a\rangle^2$. Its values of $0\pm0.0139$ and $0\pm0.0264$ for $\pi$ and $\pi/2$-pulses show an almost complete antibunching of the microwave field, demonstrating that the output is almost purely a single photon.
To further demonstrate that our source is nonclassical, in Fig.~\ref{moment}(b), we reconstruct the Wigner function from the relation $W(\alpha)=(2/{\pi})\Tr{[\hat{D}(\alpha)\rho \hat{D}^{\dag}(\alpha)\hat{\Pi}]}$, by using a maximum likelihood method~\cite{James2001,eichler2012characterizing}, where $\hat{D}(\alpha)$ is the displacement operator with a coherent state $\alpha$, $\hat{\Pi}$ is the parity operator and $\rho$ is the extracted density matrix of the filtered output from the different orders of moments.

Besides the photon leakage, there are a number of different properties that are important for proper operation of the single-photon source, such as frequency tunability, quantum efficiency, stability, bandwidth and repetition rate. In the following paragraphs we study and evaluate these quantities for our single-photon source.

{\bf{Bandwidth, repetition rate and tunability.}} The repetition rate for our source is limited by the coupling strength between the qubit and the transmission line which can be varied over a wide range by design. For our sample the coupling strength is approximately $2\pi\times 200\unite{kHz}$ resulting in a repetition time of about $1\mu s$.

Our single-photon source is frequency-tunable over a wide frequency range. The operation frequency is adjusted by changing the qubit frequency with the external magnetic flux and adjusting the frequency of the microwave source that generates the $\pi$-pulse and the cancellation pulse. Here we show tunability of up to $600\unite{MHz}$, where it is limited by flux noise producing large jumps in the qubit frequency when the qubit is tuned too far away from the flux sweet spot ($\Phi=0$).


{\bf{Intrinsic quantum efficiency.}} Different from the overall quantum efficiency, the intrinsic quantum efficiency only depends on the qubit coherence, which is the upper bound for the overall efficiency. We also investigate the intrinsic quantum efficiency which is given by $\eta_{\rm{q}}=\Gamma_{\rm{r}}/(2\Gamma_2)$, of our single photon source over the frequency range 4.9-5.5\unite{GHz}. The quantum efficiency is in the range $71 - 99\%$ [red In Fig.~\ref{efficiency}(a)], extracted from the reflection coefficient.
Typically, the pure dephasing rate $\Gamma_{\phi}$ can decohere the supposition of vacuum and a single-photon Fock state, resulting in a decrease in the single-photon quantum efficiency. Moreover, a single photon can be dissipated into the environment through the nonradiative decay channel due to the qubit interaction with the environment.
We denote that the reduction of the quantum efficiency from these two effects as $\eta_{\rm{p}}=\Gamma_{\phi}/\Gamma_{2}$ and $\eta_{\rm{n}}=\Gamma_{\rm{n}}/2\Gamma_{2}$, respectively. Here, the values of $\eta_{\rm{p}}$ are based on the exponential decay from the qubit emission as discussed below [black,in Fig.~\ref{efficiency}(a)]. Then, we calculate $\eta_{\rm{n}}$ indirectly, from $\eta_{\rm{n}}=\Gamma_{\rm{n}}/2\Gamma_{2}=1-\eta_{\rm{p}}-\eta_{\rm{q}}$ [blue, in Fig.~\ref{efficiency}(a)]. In Fig.~\ref{efficiency}(a), we find that the nonradiative decay only affects the quantum efficiency near the maximal qubit frequency the quantum efficiency. When we tune the qubit frequency down, the pure phasing dominates the reduction of the quantum efficiency. Therefore, it is necessary to understand which type of noise induces the pure dephasing rate.

To extract $\Gamma_{\phi}$, we send a pulse with the amplitude close to a $\pi$/2-pulse, and measure the qubit emission with $3.84\times 10^7$ averages. From the emission decay, we can extract both $\Gamma_{1}$ and $\Gamma_{2}$, the power decay $\propto e^{-\Gamma_{\rm{1}}t}$, and the quadrature decay $\propto e^{-\Gamma_{\rm{2}}t}$. Then, $\Gamma_{\phi}$ can be calculated from $\Gamma_{\phi}=\Gamma_{2}-\Gamma_{1}/2$. In Fig.~\ref{efficiency}(b), the data (black) shows that the pure dephasing rate increases when the qubit is tuned away from the flux sweet spot further. The averaged pure dephasing rate $\overline{\Gamma_{\phi}}$ over the whole frequency range is about $2\pi*10\unite{Hz}$. The pure dephasing rate $\Gamma_{\phi}$ due to $1/f$ flux  noise with the flux noise spectral density $S_{\Phi}(f)=A_{\Phi}/f$ has the relationship  $\Gamma_{\phi}=\sqrt{A_{\Phi}|\ln(2\pi f_{\rm{IR}}t)|}\frac{\partial \omega_{01}}{\partial\Phi}$ \cite{hutchings2017tunable}. $f_{\rm{IR}}$ is the infrared cutoff frequency, taken to be 5\unite{mHz} determined by the measurement time, and $t$ is on the order of $\overline{\Gamma_{\phi}}^{-1}$. Using this relationship to fit the extracted $\Gamma_{\phi}$ values shown as a dashed line in Fig.~\ref{efficiency}(b), we obtain $A_{\Phi}^{1/2}\approx2\unite{\mu\Phi_0}$, which is consistent with other measurements \cite{hutchings2017tunable,bialczak20071}.

\begin{figure}
\includegraphics[width=1\linewidth]{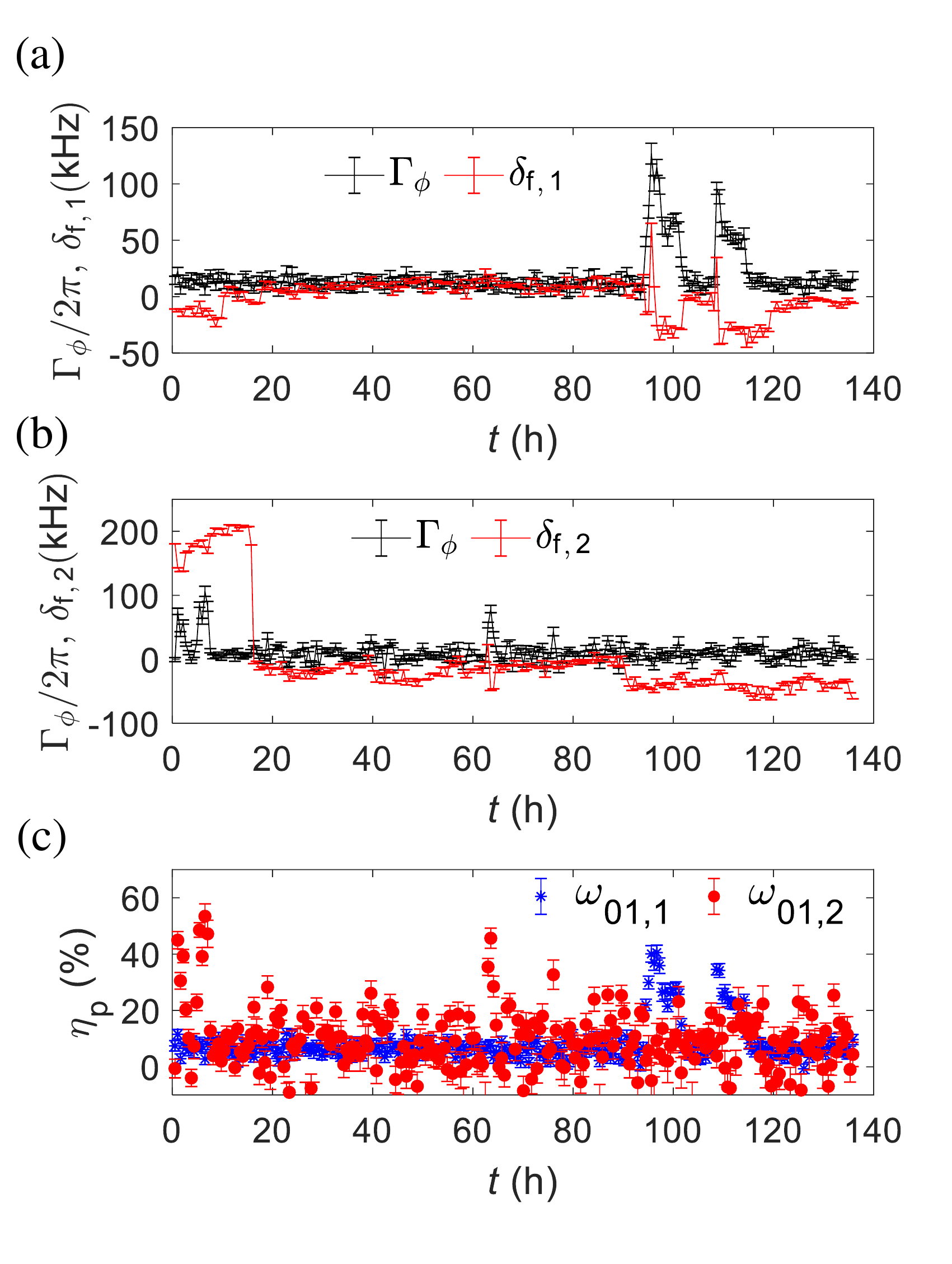}
\caption{
(a) and (b) Fluctuations on the pure dephasing rate $\Gamma_{\phi}$ and the qubit frequency around $\omega_{01,1}/(2\pi)=5.51\unite{GHz}$ and $\omega_{01,2}/(2\pi)=5.39\unite{GHz}$, corresponding to 0\unite{MHz} and 120\unite{MHz} detunings, over 136 hours.
(c) Fluctuations of the reduction of the quantum efficiency, $\eta_{\rm{p}}$, due to the fluctuations of $\Gamma_{\phi}$ at 0\unite{MHz} and 120\unite{MHz} detunings, over 136 hours.
}
\label{fluctuation}
\end{figure}

\begin{figure}
\includegraphics[width=\linewidth]{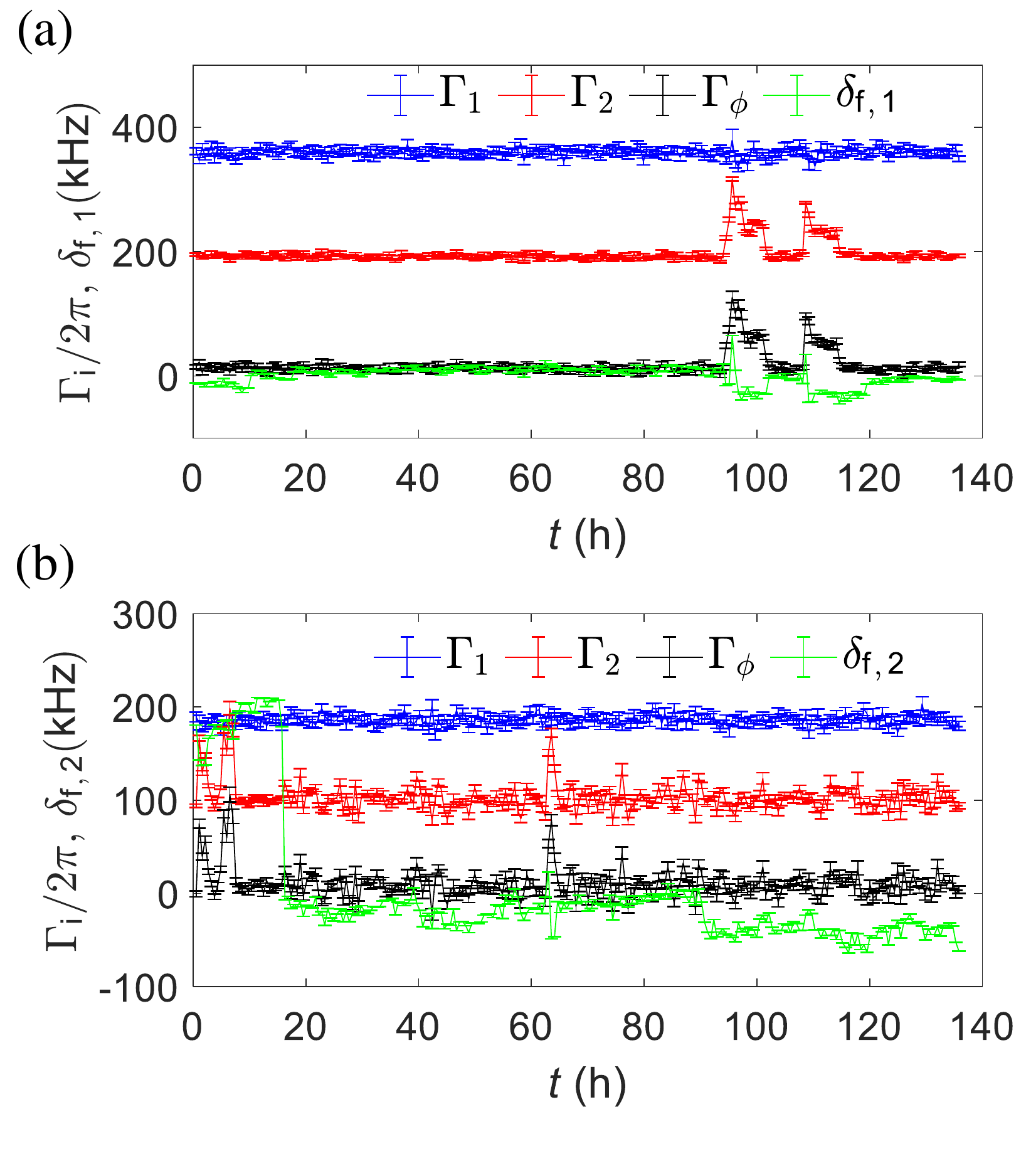}
\caption{
(a) Fluctuations on the decay rates and the qubit frequency at $\omega_{01,1}/(2\pi)=5.51\unite{GHz}$,  over 136 hours.
(b) Fluctuations on the decay rates and the qubit frequency at $\omega_{01,2}/(2\pi)=5.39\unite{GHz}$,  over 136 hours.
In all panels, the error bars are for two standard deviations.
}
\label{fluctuationGamma1Gamma2}
\end{figure}

{\bf{Stability.}} Recently, many works demonstrated that fluctuating TLSs can limit the coherence of superconducting qubits \cite{klimov2018fluctuations,schlor2019correlating,lisenfeld2019electric,burnett2019decoherence}. Here, we investigate how the fluctuations affect different properties of our single photon source. We repeatedly measure $\Gamma_{1}$ and $\Gamma_{2}$ interleaved at $\Phi=0$ and $\Phi=0.09\unite{\Phi_0}$, corresponding to $\omega_{01,1}=\omega_{01}(0)=2\pi\times5.51\unite{GHz}$ and $\omega_{01,2}=\omega_{01}(0.09\unite{\Phi_0})=2\pi\times5.39\unite{GHz}$, respectively. At the same time, the fluctuations of the qubit frequency are also obtained from the phase information of the emitted field which carries information about the qubit operator $\langle\sigma_-\rangle \propto e^{i\delta\omega_{01}t}$ where $\delta\omega_{01}$ is the frequency difference between the frequency of the driving pulse and the qubit frequency. The total measurement spans $4.90\times10^5\unite{s}$ (approximately 136 hours) with 2000 repetitions for each qubit frequency. Each repetition has $3.20\times10^6$ averages. From the values of $\Gamma_1$ and $\Gamma_2$, we extract $\Gamma_{\phi}$ values shown in Fig.~\ref{fluctuation}(a) and (b) from averaging over 8 repetitions.
We find that $\Gamma_{1}$ remains stable for both zero detuning and 120\unite{MHz} detuning in Fig.~\ref{fluctuationGamma1Gamma2}. By assuming that $\Gamma_{\rm{r}}$ is stable over time, this implies that for this detuning $\Gamma_{\rm{n}}$ is also stable on the scale of $\Gamma_{\rm{r}}$.

However, the fluctuations of the qubit frequency, $\delta_{f,i}=(\omega_{01,i}-\kb{\omega_{01,i}})/2\pi$ and the pure dephasing rate are obvious as shown in Fig.~\ref{fluctuation}(a) for $\delta_{f,1}$ and (b) for $\delta_{f,2}$. First, we note the frequency jumps for the case of 120\unite{MHz} detuning (i.e. around $\omega_{01,2}$) at $t=16\unite{h}$ and $t=90\unite{h}$ do not affect the pure dephasing rate. We suspect that this is due to a change in the flux offset through the SQUID, as we tune the qubit back and forth by the applied external flux that could induce a change in magnetic polarization in cold components. Therefore, we can not see significant fluctuations at the flux sweet spot.

Other frequency-switching events happening at $t=95\unite{h}$ and $t=120\unite{h}$ for 0\unite{MHz} detuning (i.e. around $\omega_{01,1}$) and those before $t=10\unite{h}$ and at $t=64\unite{h}$ for 120\unite{MHz} detuning show a strong positive correlation with the pure dephasing rate. Interestingly, the fluctuations do not happen at the same time for both detunings. Combining this with the fact that $\Gamma_1$ is stable, we speculate that this is due to two uncorrelated TLSs with a small decay rate $\gamma_{\rm{i}}$ (i=1,2), close to $\omega_{01,\rm{i}}$, dispersively coupled to the qubit [see more details in the section Methods]. Thus, these two TLSs can only cause the pure dephasing, but not dominate the relaxation, which can explain the stronger fluctuations in $\Gamma_{\phi}$ compared to $\Gamma_1$ shown in Fig.~\ref{fluctuationGamma1Gamma2}.

Evidently,  these two TLSs reduce the intrinsic quantum efficiency substantially by up to $40\%$ and $60\%$ as shown in Fig.~\ref{fluctuation}(c) for detunings of 0\unite{MHz} and 120\unite{MHz}, respectively. The effect from TLSs is stronger than other types of noises, especially in the case of zero detuning. At zero detuning we also note that between these large fluctuations the single-photon source can be stable for tens of hours. However, the qubit becomes more sensitive to the $1/f$ flux noise when it is detuned by 120\unite{MHz}, it results in about a $20\%$ fluctuation of the quantum efficiency over the total measurement time. This indicates that $1/f$ flux noise will be the dominant noise when we tune the qubit frequency away from the flux sweet spot.

Since our single-photon source has a narrow bandwidth it will be meaningful to investigate the frequency stability over a long time, from Fig.~\ref{fluctuation}(a) and (b), we find that at $\Phi=0$ the frequency fluctuations due to TLSs can be up to 100\unite{kHz} which is nearly one third of the single-photon linewidth ($\Gamma_{r}=270\unite{kHz}$). However, just tuned down the qubit frequency by 120\unite{MHz} ($\Phi=0.09$), the external flux jumps described above dominate the frequency shift of the single-photon source, the shifts can be up to $200\unite{kHz}$ which is a factor of two compared to the effect from TLSs.

\section{Discussion}
\label{Discussion}
In this paper, we demonstrate a method to implement a frequency-tunable single photon source by using a superconducting qubit. We measure the moments of the emitted field, and from those we can evaluate both the second order correlation function and the Wigner function. Our study illustrates that the intrinsic quantum efficiency of our single-photon source can reach up to $99\%$, which could be improved further by engineering a large radiative decay rate of the qubit into the waveguide transmission line. Moreover, the photon leakage from the cancelled input $\pi$-pulse is as low as 0.5\% of a photon, indicating that our single-photon source is very pure. The frequency tunable range of our single photon source corresponds to $1600\times\Gamma_{1}$, reaching state of the art and enabling us to address quantum memories with a large number of different `colors.'

We also study the noise mechanisms which limit the intrinsic quantum efficiency in detail. The nonradiative decay rate and the pure dephasing rate from the $1/f$ flux noise both contribute to the reduced quantum efficiency. The $1/f$ flux noise could be decreased by reducing the density of surface spins by surface treatment of the sample, e.g.\@ annealing~\cite{de2018suppression} and UV illumination~\cite{kumar2016origin}.

Finally, we investigate the stability of our single photon source, which is important for long time operation. The instability originates mainly from the increased sensitivity to $1/f$ flux noise when the source frequency is tuned down from the flux-insensitive bias point. The results show that the source can be stable for tens of hours at the maximum frequency. However, sometimes, the quantum efficiency decreases by up to $60\%$ when the qubit couples to TLSs. Besides reducing the quantum efficiency, the TLSs can also change the frequency of the single photons by up to one third of the linewidth. However, the environment flux jump will be the dominant noise to shift the single-photon frequency, which could be further reduced by magetic shields e.g. Cryoperm shielding~\cite{burnett2019decoherence,kreikebaum2016optimization}.

\begin{figure}
\includegraphics[width=\linewidth]{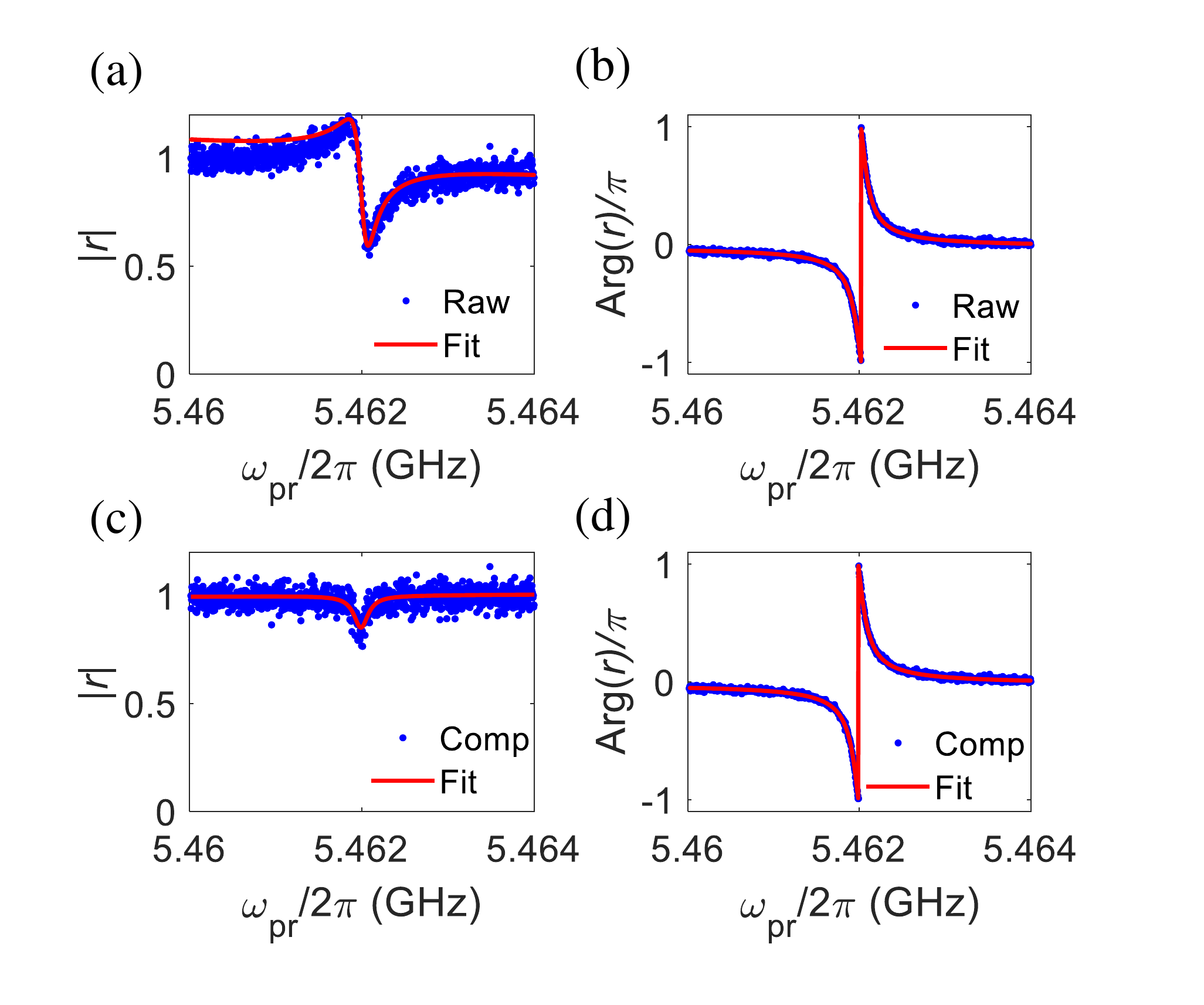}
\caption{Impedance mismatch. (a) and (b) are the magnitude and phase response of the reflection coefficient before compensating the impedance mismatch. (c) and (d) are the magnitude and phase response of the reflection coefficient after compensating the impedance mismatch.
}
\label{impedance}
\end{figure}
\section{Methods}
{\bf{Measurement setup and qubit characterization.}}
Figure.~\ref{setup}(a) shows the detailed experimental setup. To characterize the qubit, a vector network analyzer (VNA) generates a weak coherent probe with the frequency $\omega_{\rm{pr}}$. The signal is fed into the input line, attenuated to be weak ($\Omega<\Gamma_1$) and interacts with the qubit. Then, the VNA receives the reflected signal from the output line after the amplification to determine the complex reflection coefficient, $r$. Two-tone spectroscopy is then done to obtain the qubit anharmonicity. Specifically, we apply a strong pump at $\omega_{01}$ to saturate the $|0\rangle-|1\rangle$ transition. Meanwhile, we combine a weak probe with the strong pump together via a 20 dB directional coupler. The frequency of the weak probe from the VNA is swept near the $|1\rangle-|2\rangle$ transition. When the probe is on resonance, again, we will get a dip in the magnitude response of $r$, leading to $\alpha=(\omega_{01}-\omega_{12})/\hbar=2\pi*0.251\unite{GHz}$ (not shown).

\begin{figure}
\includegraphics[width=\linewidth]{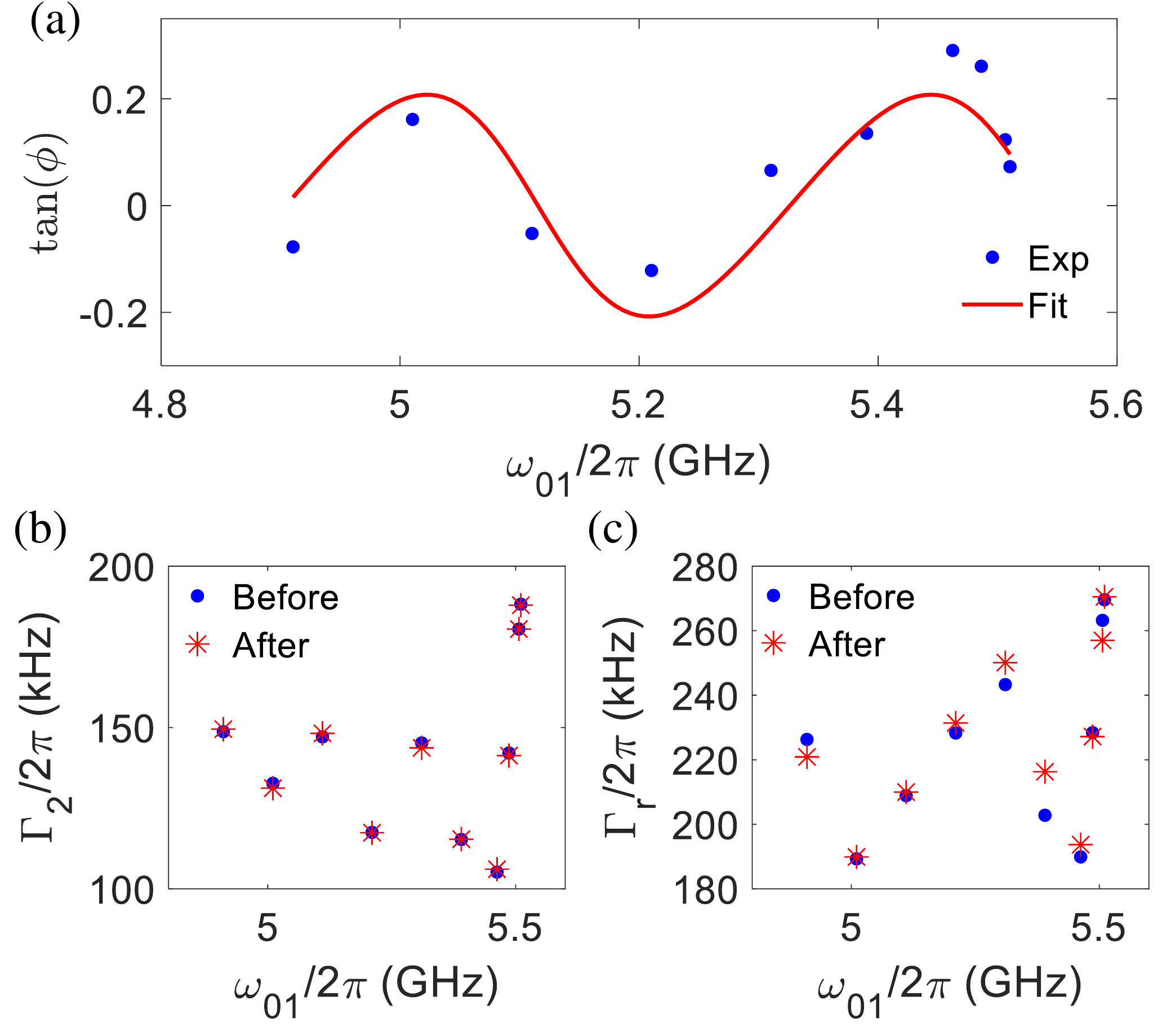}
\caption{ Compensating the Fano-shape reflection coefficient. (a) The phase $\phi$ in Eq. \ref{phase} vs. different qubit frequencies. Blue dots are the data with the solid curve as a fit. (b) and (c) The comparison of $\Gamma_2$ and $\Gamma_{\rm{r}}$ before and after compensating the impedance mismatch.
}
\label{phi}
\end{figure}

{\bf{Fano-shape spectroscopy.}} When we measure the reflection coefficient at different qubit frequencies, we notice that at some frequencies the amplitude of the spectroscopy is not flat but has a Fano shape [Fig.~\ref{impedance}(a)]. This Fano shape may affect the extracted $\Gamma_{\rm{r}}$ values, and we argue that the Fano shape originates from an impedance mismatch in the measurement setup which will result in a modified reflection coefficient as \cite{lu2021propagating}:
\begin{equation}
r = 1-\frac{i\Gamma_{\rm{r}}e^{i\phi}}{\Delta + i\Gamma_2},
\label{reflection}
\end{equation}
where
\begin{equation}
\tan(\phi) = \frac{r_1\sin2\phi_0}{t_1^2\beta^2+r_1\cos2\phi_0},
\label{phase}
\end{equation}
$r_1$ ($t_1$) is the reflection (transmission) coefficient at the place where the impedance mismatch is located, and $\beta$ is proportional to the attenuation between the place and the sample. $\phi_0=\omega\tau$ is the extra phase of the propagating wave from the propagating time $\tau$, due to the distance between the qubit and the impedance mismatch. We use Eq.~(\ref{reflection}) to fit the data to extract the values of $\Phi$ at different qubit frequencies which are thus fit to Eq.~(\ref{phase}) as show in Fig~\ref{phi}(a). The extracted $r_1\approx0.14$ close to 0.1 (corresponding -20\unite{dB} in power) and $\beta\approx0.97$ corresponding to 0.26 dB attenuation indicate that the impedance mismatch probably arises from the directional coupler.

Afterwards, to compensate the impedance mismatch, we calculate $r_{\rm{comp}}=1-(1-r_{\rm{raw}})*e^{i\phi}$ where $r_{\rm{raw}}$ is the raw data [blue in Fig~\ref{impedance}(a) and (b)]. The magnitude response of $r_{\rm{comp}}$ in Fig~\ref{impedance}(c) manifests that the impedance mismatch has been corrected. We repeat this process for other qubit frequencies and then fit the calculated data to obtain $\Gamma_{\rm{r}}$ and $\Gamma_{\rm{2}}$ [red stars in Fig~\ref{phi}(b) and (c)]. Comparing to the values before correcting the impedance mismatch [blue dots in Fig~\ref{phi}(b) and (c)], we find they are close to each other.

{\bf{Two-level fluctuator model.}}
Figure~\ref{fluctuationGamma1Gamma2} shows the fluctuations of $\Gamma_1$ and $\Gamma_2$ at $\omega_{01,1}/(2\pi)=5.51\unite{GHz}$ and $\omega_{01,2}/(2\pi)5.39\unite{GHz}$ over 136 hours.
We denote $g_{\rm{i}}$ and $\Delta_{\rm{i}}=(\omega_{\rm{TLS,i}}-\omega_{01,\rm{i}})/(2\pi)$ as the coupling strength and the frequency detuning between the TLS and the qubit, respectively. In addition, $\Gamma_{\rm{n},\rm{i}}$ and $\Gamma_{\rm{1},\rm{i}}$ are the corresponding non-radiative decay rate and relaxation at each qubit frequency. To simplify the model, we let $g_{\rm{1}}=g_{\rm{2}}$.  When $g_{\rm{i}}\ll\Delta_{\rm{i}}\ll120\unite{MHz}$, we have a dispersive shift $\chi_{\rm{i}}=g_{\rm{i}}^2/\Delta_{\rm{i}}$. Typically, the surface TLS coupling rates are on the order of $g\approx 100\unite{kHz}$~\cite{schlor2019correlating}. Since the measured frequency shifts of both qubit frequencies are almost the same, about 40\unite{kHz}, the detuning to such a TLS is approximately $\Delta_{\rm{i}}=g_{\rm{i}}^2/\chi_{\rm{i}}=2.5~\rm{MHz}$, which is about $9\times\Gamma_{\rm{1},\rm{i}}$. From the shortest duration of the TLS fluctuations in Fig.~\ref{fluctuation}(b), we can estimate the switching time of these two TLSs roughly to be $2.88\times10^4\unite{s}$ and $7.82\times10^3\unite{s}$, corresponding to $\gamma_{\rm{1}}=34.7\unite{\mu Hz}$ and $\gamma_{\rm{2}}=127.9\unite{\mu Hz}$, respectively. According to $\Gamma_{\rm{n},\rm{i}}\propto g_{\rm{i}}^2/\Delta_{\rm{i}}^2\gamma_{\rm{i}}=0.16\%\gamma_{\rm{i}}$. Thus, these two TLSs can only cause the pure dephasing, but not dominate the relaxation. This can also explain the stronger fluctuations in $\Gamma_{\phi}$ compared to $\Gamma_1$ shown in Fig.~\ref{fluctuationGamma1Gamma2}. We emphasize that the fresh finding here is that we notice TLSs can be activated independently where there is only a single TLS was investigated in Ref.~\cite{schlor2019correlating}.

\section{Data availability}
The data that supports the findings of this study is available from the corresponding authors upon reasonable request.
\section*{Code availability}
The code that supports the findings of this study is available from the corresponding authors upon reasonable request
\section*{Acknowledgements}
The authors acknowledge the use of the Nano fabrication Laboratory (NFL) at Chalmers. We also acknowledge IARPA and Lincoln Labs for providing the TWPA used in this experiment.
We wish to express our gratitude to Lars J\"{o}nsson for help and we appreciate the fruitful discussions with Prof. Simone Gasparinetti, Dr. Marek Pechal, Dr. Neill Lambert and  Ingrid Strandberg. This work was supported by the Knut and Alice Wallenberg Foundation via the Wallenberg Center for Quantum Technology (WACQT) and by the Swedish Research Council. B.S. acknowledges the support of the DST-SERB-CRG and Infosys Young Investigator grants.
\section*{AUTHOR CONTRIBUTIONS}
P.D. and Y.L. planned the project. Y.L. performed the measurements with the input from A.B., J.J.B, B.S., and P.D. . Y.L. designed and A.B. fabricated the sample. Y.L. developed the theoretical expressions. H.R.N. and Y.L. set up the TWPA. Y.L. wrote the manuscript with input from all the authors. Y.L. analyzed the data with inputs from J.B. and P.D..
 P.D. supervised this work.

\section*{REFEREENCE}
\bibliographystyle{naturemag}


\makeatletter
\renewcommand{\fnum@figure}{\figurename~S\thefigure}
\makeatother
\renewcommand{\thefigure}{S\arabic{figure}}
\renewcommand{\thesection}{S\arabic{section}}
\renewcommand{\theequation}{S\arabic{equation}}

\end{document}